\documentclass[twoside]{article}
\usepackage{fleqn,espcrc2,epsf}
\usepackage{graphicx}
\usepackage{epsfig}
\usepackage[figuresright]{rotating}

\newcommand{\be}[1]{\begin{equation} \label{(#1)}}
\newcommand{\ee}{\end{equation}}
\newcommand{\ba}[1]{\begin{eqnarray} \label{(#1)}}
\newcommand{\ea}{\end{eqnarray}}

\newcommand{\AmS}{{\protect\the\textfont2
  A\kern-.1667em\lower.5ex\hbox{M}\kern-.125emS}}

\def\be{\begin{equation}}
\def\ee{\end{equation}}
\def\bea{\begin{eqnarray}}
\def\eea{\end{eqnarray}}
\def \KK {H.V.~Klapdor-Kleingrothaus}
\title{Neutrino Mass from Laboratory:\\
Contribution of Double Beta Decay to the Neutrino Mass Matrix}

\author{H.V. Klapdor--Kleingrothaus
\address{Max--Planck--Institut f\"ur Kernphysik, 
P.O.Box 10 39 80, D--69029 Heidelberg, Germany\\
Spokesman HEIDELBERG-MOSCOW and GENIUS Collaborations\\
e-mail:klapdor@gustav.mpi-hd.mpg.de, home page: 
http://mpi-hd.mpg.de.non$\_$acc/}
}

\begin{document}

\begin{abstract}Double beta decay is indispensable to solve the question 
	of the neutrino mass matrix together with $\nu$ oscillation 
	experiments. 
	The most sensitive experiment - since eight years the 
	HEIDELBERG-MOSCOW experiment in Gran-Sasso - already now, with the 
	experimental limit of $\langle m_\nu \rangle < 0.26$ eV practically 
	excludes 
	degenerate $\nu$ mass scenarios allowing neutrinos as hot dark 
	matter in the universe for the smallangle MSW solution of the solar 
	neutrino problem. It probes cosmological models including hot 
	dark matter already now on the level of future satellite experiments 
	MAP and PLANCK. It further probes many topics of beyond SM physics 
	at the TeV scale. Future experiments should 
	give access to the multi-TeV range and complement on many ways 
	the search for new physics at future colliders like LHC and NLC. 
	For neutrino physics some of them (GENIUS) will allow to test 
	almost {\it all} neutrino mass scenarios 
	allowed by the present neutrino oscillation experiments.

\vspace{1pc}
\end{abstract}

\maketitle

\section{Introduction}

	Recently atmospheric and solar neutrino oscillation experiments 
	have shown that neutrinos are massive. This is the first 
	indication of beyond standard model physics. The absolute 
	neutrino mass scale is, however, still unknown, 
	and only neutrino oscillations and neutrinoless double beta decay 
	{\it together} can solve this problem (see, e.g. 
\cite{KKPS,KKP,KK60Y}).

	In this paper we will discuss the contribution, that can be 
	given by present and future $0\nu\beta\beta$ experiments to this 
	important question of particle physics. We shall, in section 2, 
	discuss the expectations for the observable of neutrinoless double 
	beta decay, the effective neutrino mass $\langle m_\nu \rangle$, 
	from the most recent $\nu$ oscillation experiments, 
	which gives us the required sensitivity for future $0\nu\beta\beta$ 
	experiments. In section 3 we shall discuss the present status 
	and future potential of $0\nu\beta\beta$ experiments. 
	It will be shown, that if by exploiting the potential of 
	$0\nu\beta\beta$ decay to its ultimate experimental limit, it will 
	be possible to test practically 
	{\it all} neutrino mass scenarios allowed by the present neutrino 
	oscillation experiments (except for one, the hierarchical 
	LOW solution). 


\section{Allowed ranges of $\langle m \rangle$ by $\nu$ oscillation 
experiments}

	After the recent results from Superkamiokande (e.g. see 
\cite{Gonz00}), the prospects for a positive signal in 
	$0\nu\beta\beta$ decay have become more promising.
	The observable of double beta decay  
$\langle m \rangle = |\sum U_{ei}^2 m_i| = 
|m^{(1)}_{ee}| + e^{i\phi_{2}} |m_{ee}^{(2)}|
+  e^{i\phi_{3}} |m_{ee}^{(3)}|$
	with $U_{ei}$ denoting elements of the neutrino mixing matrix, 
	$m_i$ neutrino mass eigenstates, and $\phi_i$ relative 
	Majorana CP phases, can be written in terms of oscillation 
	parameters 
\cite{KKPS,KKP}
\be{}
|m^{(1)}_{ee}|  ~=~  |U_{e1}|^2 m_1, 
\ee
\be{}
|m^{(2)}_{ee}|~=~|U_{e2}|^2 \sqrt{\Delta m^2_{21} + m_1^2},
\ee
\be{}
|m^{(3)}_{ee}|~=~|U_{e3}|^2\sqrt{\Delta m^2_{32}+ \Delta m^2_{21} + m_1^2}.
\label{gg}
\ee

	The effective mass $\langle m \rangle$ is related with the half-life 
	for $0\nu\beta\beta$ decay via 
${(T_{1/2}^{0\nu})}^{-1} \sim {\langle m_\nu \rangle}^2$, and for the limit on 
$T_{1/2}^{0\nu}$ deducable in an experiment we have 
$T_{1/2}^{0\nu}\sim a \sqrt{\frac{M t}{\Delta E B}}$.
	Here are a - isotopical abundance of the $\beta\beta$ emitter;  
	M - active detector mass; t - measuring time; 
	$\Delta E$ - energy resolution; B - background count rate.
	Neutrino oscillation experiments fix or restrict some of the 
	parameters in eqs. 1-3, e.g. in the case of normal hierarchy 
	solar neutrino experiments yield $\Delta m^2_{21}$, 
	$|U_{e1}|^2=\cos^2 \theta_{\odot}$ and 
	$|U_{e2}|^2=\sin^2 \theta_{\odot}$. Atmospheric neutrinos fix 
	$\Delta m^2_{32}$ and experiments like CHOOZ, looking for 
	$\nu_e$ disapperance restrict $|U_{e3}|^2$. The phases $\phi_i$ and 
	the mass of the lighest neutrino, $m_1$ are free parameters. 
	The expectations for $\langle m \rangle$ from oscillation 
	experiments in different neutrino mass scenarios have been 
	carefully analyzed in 
\cite{KKPS,KKP}.

\vspace{-.7cm}
\begin{figure}[htb]
\vspace{9pt}
\centering{
\includegraphics[width=0.45\textwidth]{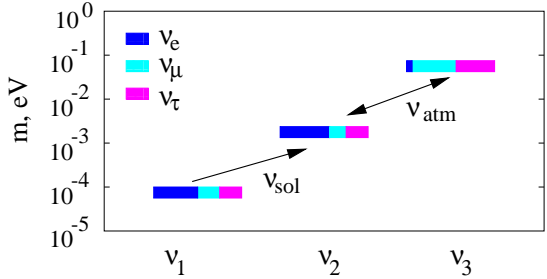}}
\vspace{-.7cm}
\caption[]{Neutrino masses and mixings in the scheme with mass hierarchy. 
	Coloured bars correspond to flavor 
	admixtures in the mass eigenstates $\nu_1$, $\nu_2$, $\nu_3$. 
	The quantity $\langle m \rangle$ is determined by the dark blue
	 bars denoting the admixture of the electron neutrino $U_{ei}$.
\label{smi1}}
\end{figure}


\vspace{-.7cm}
\subsection{Hierarchical spectrum ($m_1\ll m_2\ll m_3$)}
	
\begin{figure}[!ht]
\begin{center}
\includegraphics[width=0.45\textwidth]{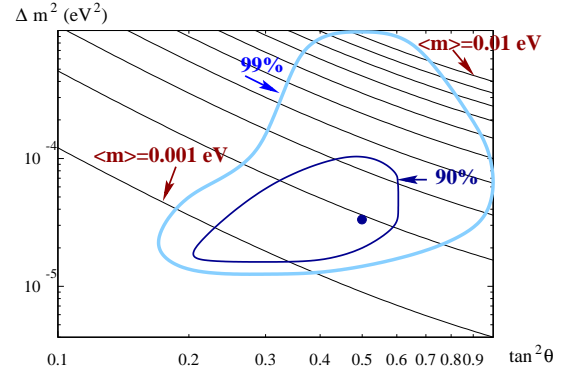}
\end{center}
\vspace{-.7cm}
\caption[]{
	Double beta decay observable $\langle m \rangle$ and oscillation 
	parameters 
	in the case of the MSW large mixing solution of the solar 
	neutrino deficit, where the dominant contribution to 
$\langle m \rangle$
	comes from the second state. Shown are lines 
	of constant $\langle m \rangle$, the lowest line corresponding to 
	$\langle m_\nu \rangle$ = 0.001 eV, the upper line to 0.01 eV.
	The inner and outer closed line show the regions allowed by present 
	solar neutrino experiments with 90 \% C.L. and 99 \% C.L., 
	respectively. 
	Double beta decay with sufficient sensitivity could check the LMA 
	MSW solution.
	Complementary information could be obtained
	from the search for
	a day-night effect and spectral distortions 
	in future solar neutrino experiments as well as a disappearance 
	signal in KAMLAND. 
\label{dark2}}
\end{figure}

	In hierarchical spectra (Fig. \ref{smi1}), motivated by analogies 
	with the quark sector and the simplest 
	see-saw models, the main contribution comes from $m_2$ or $m_3$. 
	For the large mixing angle (LMA) MSW solution which is favored 
	at present for the solar neutrino problem (see 
\cite{Suz-Neutr2000}), the contribution of $m_2$ becomes dominant 
	in the expression for $\langle m \rangle$, and 
\be{}
\langle m \rangle \simeq m_{ee}^{(2)}= 
\frac{\tan^2 \theta}{1+ \tan^2 \theta} \sqrt{\Delta m_{\odot}^2}.
\ee
	In the region allowed at 90\% c.l. by Superkamiokande according to 
\cite{Gonz00} the prediction for $\langle m \rangle$ becomes 
\be{}
\langle m \rangle = (1-3) \cdot 10^{-3}~{\rm eV}.
\ee
	The prediction extends to $\langle m \rangle=10^{-2}$ eV in 
	the 99\% c.l. range (Fig. \ref{dark2}).


\vspace{-.9cm}
\subsection{Inverse Hierarchy ($m_3 \approx  m_2 \gg m_1$)}

\begin{figure}[htb]
\vspace{9pt}
\centering{
\includegraphics[width=0.45\textwidth]{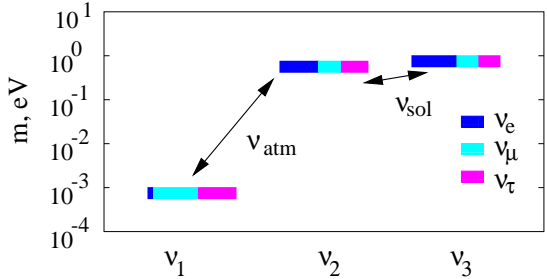}}
\vspace{-0.6cm}
\caption[]{Neutrino masses and mixings in the inverse hierarchy scenario.
\label{smi2}}
\end{figure}

	In inverse hierarchy scenarios (Fig. \ref{smi2}) the heaviest 
	state with 
	mass $m_3$ is mainly the electron neutrino, its mass being 
	determined by atmospheric neutrinos,  
$m_3 \simeq \sqrt{\Delta m_{atm}^2}$. 
	For the LMA MSW solution one finds
\cite{KKP}
\be{}
\langle m \rangle = (1-7) \cdot 10^{-2}~{\rm eV}.
\ee 

\vspace{-.1cm}
\subsection{Degenerate spectrum 
($m_1 \simeq  m_2 \simeq  m_3 \ge\sim 0.1 eV$)}

	Since the contribution of $m_3$ is strongly restricted by CHOOZ, 
	the main contributions come from $m_1$ and $m_2$, depending on 
	their admixture to the electron flavors, which is determined 
	by the solar neutrino solution. We find \cite{KKP} 
\be{}
m_{min} < \langle m \rangle < m_1~~~~with
\label{m-min}
\ee 
\ba{}
\langle m \rangle_{min} = (\cos^2 \theta_{\odot}-\sin^2 
\theta_{\odot})~m_1 \nonumber.
\ea

	This leads for the LMA solution to 
$\langle m \rangle = (0.25 - 1) \cdot m_1$, the allowed range corresponding 
	to possible values of the unknown Majorana CP-phases.

\begin{figure}[htb]
\vspace{9pt}
\centering{
\includegraphics[width=0.5\textwidth]{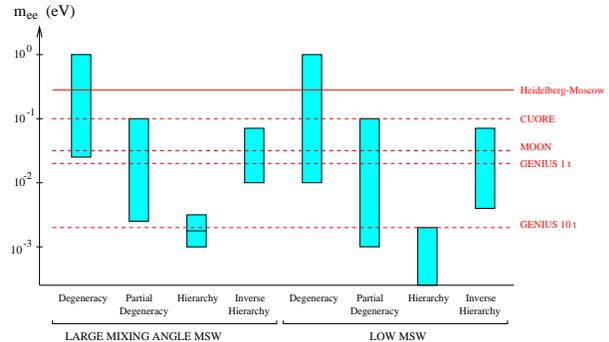}}
\vspace{-0.7cm}
\caption[]{Summary of values for $m_{ee} \equiv \langle m \rangle $ 
	expected from neutrino oscillation 
	experiments (status NEUTRINO2000), in the different 
	schemes discussed in this paper. For a more general analysis see 
\cite{KKPS}.  
	The expectations are compared with the recent neutrino mass limits 
	obtained from the HEIDELBERG-MOSCOW 
\cite{hdmo,KK-AnnRep00},
	experiment as well as the expected sensitivities for the CUORE 
\cite{cuore}
, MOON 
\cite{moon}, EXO 
\cite{exo} proposals and the 1 ton and 10 ton proposal of GENIUS 
\cite{KK-BEY97,KKPropos99}.
\label{NewSm-Pfig}}
\end{figure}

	After these examples we give a summary of our analysis 
\cite{KKPS,KKP} of the $\langle m \rangle$ allowed by $\nu$ oscillation 
	experiments for the neutrino mass models in the presently 
	favored scenarios, in 
Fig. \ref{NewSm-Pfig}. 
	The size of the bars corresponds to the uncertainty in mixing 
	angles and the unknown Majorana CP-phases.

\section{Status and Future of $\beta\beta$ Experiments}

\begin{figure}[htb]
\vspace{9pt}
\centering{
\includegraphics[width=0.35\textwidth, angle=-90]{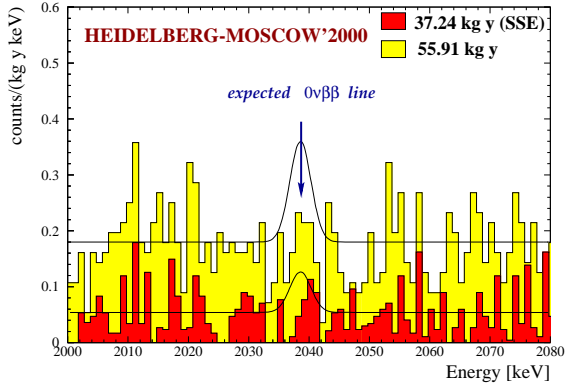}}
\vspace{-0.7cm}
\caption[]{HEIDELBERG-MOSCOW experiment: energy spectrum in the range 
	between 2000 keV and 2080 keV, where the peak from neutrinoless 
	double beta decay is expected. The open histogram denoteds the 
	overal sum spectrum without PSA after 55.9 kg y of measurement 
	(since 1992). The filled histogram corresponds to the SSe data 
	after 37.2 kg y. Shown are also the excluded (90\%) peak areas 
	from the two spectra.} 
\label{Spectr2000}
\end{figure}

	The status of present double beta experiments is shown in 
Fig. 1 of 
\cite{KK1-NOW00} and extensively discussed in 
\cite{KK60Y}. The HEIDELBERG-MOSCOW experiment using the largest source 
strength of 11 kg of enriched $^{76}{Ge}$ in form of five HP Ge-detectors 
	in the Gran-Sasso underground laboratory 
\cite{KK60Y}, yields after a time of 37.2 kg y of measurement 
(Fig. \ref{Spectr2000}) 
	a half-life limit of \cite{KK-AnnRep00}

\vspace{.2cm}
$T_{1/2}^{0\nu} > 2.1 (3.5) \cdot {10}^{25}~ y, ~~~~ 90\% (68\%) c.l.$

\vspace{.2cm}
\noindent
and a limit for the effective neutrino mass of 

$\langle m \rangle < 0.34 (0.26) ~eV, ~~~~ 90\% (68\%) c.l.. $


\begin{figure}[htb]
\vspace{9pt}
\centering{
\includegraphics[width=0.38\textwidth,angle=-90]{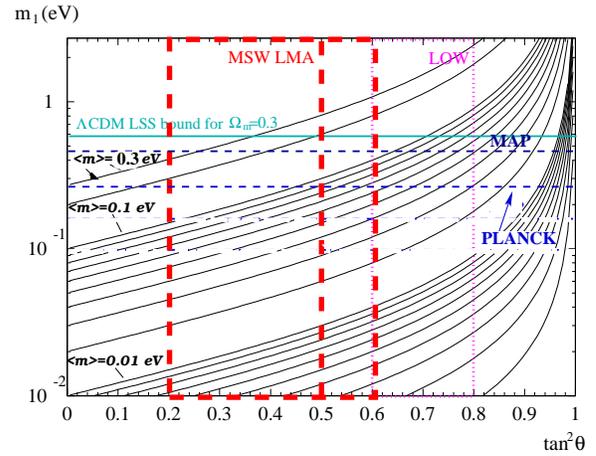}}
\vspace{-.5cm}
\caption[]{Double beta decay observable $\langle m \rangle$ and 
	oscillations parameters:
	The case for degenerate neutrinos. Plotted on the axes are 
	the overall scale of neutrino masses $m_0$
	and the mixing $\tan^2 2 \theta_{12}$.
	Also shown is a cosmological bound deduced from a fit of CMB and 
	large scale structure 
\cite{cmb} 
	and the expected 
	sensitivity of the satellite experiments MAP and Planck. 
	The present limit from tritium $\beta$ decay of 2.2 eV 
\cite{Weinh2000}
	would lie near the top of the figure. 
	The range of $\langle m \rangle$  investigated at present by the 
	HEIDELBERG-MOSCOW experiment is, in the case of small solar neutrino
	mixing already in 
	the range to be explored by MAP and Planck \protect{\cite{cmb}.}
\label{osc-param}}
\end{figure}

\vspace{0.1cm}
	This sensitivity just starts to probe some (degenerate) neutrino 
	mass models. In degenerate models from the experimental limit on 
	$\langle m \rangle$ we can conclude on upper bound on the mass 
	scale of the 
	heaviest neutrino. For the LMA solar solution we obtain from eq. 
(\ref{m-min}) $m_{1,2,3} < 1.1 eV$ implying $\sum m_i < 3.2 eV$.
	This first number is sharper than what has recently been deduced 
	from single beta decay of tritium ($m < $ 2.2 eV 
\cite{WeinLob-Neutr2000}), 
	and the second is sharper than the limit of 
	$\sum m_i <$ 5.5. eV still compatible with most recent fits of 
	Cosmic Microwave Background Radiation and Large Scale Structure 
	data (see, e.g. 
\cite{Teg-0008145}).
	The result has found a large resonance, and it has 
	been shown that it excludes for example the small angle 
	MSW solution of the solar neutrino problem in degenerate scenarios, 
	if neutrinos are considered as hot dark matter in the universe 
\cite{Glash,Min97,Yas-Bey00,Ell99}. 
Fig. \ref{osc-param} shows that the present 
	sensitivity probes cosmological 
	models including hot dark matter already now on a level of future 
	satellite experiments MAP and PLANCK. The HEIDELBERG-MOSCOW 
	experiment yields the by far sharpest limits worldwide. 
	If future searches will show that $\langle m \rangle >$ 0.1 eV, than 
	the three-$\nu$ mass schemes, which will survive, are those with 
	$\nu$ mass degeneracy or 4-neutrino schemes with inverse mass 
	hierarchy ( 
Fig. \ref{NewSm-Pfig} and \cite{KKPS}).
	It has been discussed in detail earlier (see e.g. 
\cite{KK-BEY97,KK99nu98,KK1-NOW00} 
\cite{KK60Y}), that of present generation experiments no one 
	(including NEMO-III,
	...) has a potential to probe 
	$\langle m_\nu \rangle$ below the present HEIDELBERG-MOSCOW level. 

	A possibility to probe $\langle m \rangle$ down to $\sim 0.1$ eV 
	(90\% c.l.) exists with the GENIUS Test Facility 
\cite{KKAnnRep99-00} 
	which should reduce the background by a factor of 30  compared to 
	the HEIDELBERG-MOSCOW experiment, and thus 
	could reach a half-life limit of $1.5 \cdot {10}^{26}$ y.

	To extend the sensitivity of $\beta\beta$ experiments below this 
	limit requires completely new experimental approaches, as discussed 
	extensively in 
\cite{KK-BEY97,KKPropos99,KK99nu98}, and in another contribution to this 
	conference 
\cite{KK1-NOW00}.

	Fig. \ref{NewSm-Pfig} shows that an improvement of the sensitivity 
	down to 
	$\langle m \rangle \sim {10}^{-3}$ eV is required to probe all 
	neutrino mass scenarios allowed by present neutrino oscillation 
	experiments. With this result of $\nu$ oscillation experiments nature 
	seems to be generous to us since such a sensitivity seems to be 
	achievable in future $\beta\beta$ experiment, if this method is 
	exploited to its ultimate limit (see \cite{KK1-NOW00}).

\begin{thebibliography}{9}

\bibitem{KKPS} H.V. Klapdor-Kleingrothaus, H. P\"as and A.Yu. Smirnov,
	Preprint: {\it hep-ph/}{\bf 0003219}, (2000) and 
	in {\it Phys. Rev.} {\bf D} (2000).

\bibitem{KKP} H.V. Klapdor-Kleingrothaus, H. P\"as and A.Yu. Smirnov, 
	in Proc. of DARK2000, 
	Heidelberg, 10-15 July, 2000, Germany, 
	ed H. V. Klapdor-Kleingrothaus, Springer, Heidelberg (2001).	

\bibitem{KK60Y} \KK, {\sf "60 Years of Double Beta Decay"},
	{\it World Scientific, Singapore} (2001) 1253p.

\bibitem{KKPcomm} \KK ~and H. P\"as, Preprint: {\it physics/}{\bf 0006024} 
	and {\it Comm. in Nucl. and Part. Phys.} (2000).

\bibitem{LowNu2} \KK~, in Proc. International Workshop 
	LowNu2, December 4 and 5 (2000) Tokyo, Japan, 
	ed: Y. Suzuki, {\it World Scientific, Singapore} (2001).

\bibitem{Bau-KK} L. Baudis and H.V. Klapdor-Kleingrothaus, 
	{\it Eur. Phys. J.} {\bf A 5} (1999) 441-443. 

\bibitem{hdmo}
H.V. Klapdor-Kleingrothaus et al., to be publ. 2000 and
$http://www.mpi-hd.mpg.de/non_acc/main.html$

\bibitem{cuore}
E. Fiorini et al., {\it Phys. Rep.} {\bf 307} (1998) 309. 

%
\bibitem{moon}
H. Ejiri et al., {\it nucl-ex/}{\bf 9911008}.

\bibitem{exo}
M. Danilov et al., {\it Phys. Lett.} {\bf B 480} (2000) 12-18.

\bibitem{KK-BEY97} \KK ~in Proceedings of BEYOND'97
	Germany, 8-14 June 1997, edited by  H.V. Klapdor-Kleingrothaus 
	and H.P\"as, {\it IOP Bristol} (1998) 485-531 and
	{\it Int. J. Mod. Phys.} {\bf A 13} (1998) 3953, and
	{\it J. Phys.} {\bf G 24} (1998) 483 - 516. 

\bibitem{KKPropos99} H.V. Klapdor-Kleingrothaus et al. 
	{\it MPI-Report} {\bf MPI-H-V26-1999} 
	and Preprint: {\it hep-ph/}{\bf 9910205} and in Proceedings of 
	BEYOND'99, Castle Ringberg, Germany, 6-12 June 
	1999, edited by \KK~ and I.V. Krivosheina, {\it IOP Bristol}, 
	(2000) 915 - 1014. 

\bibitem{KK99nu98} \KK, ~in Proc. of 
	(NEUTRINO 98), Takayama, Japan, 
	4-9 Jun 1998, (eds) Y. Suzuki et al. 
	{\it Nucl. Phys. Proc. Suppl.} {\bf 77} (1999) 357 - 368.

%
%
\bibitem{cmb} R.E. Lopez, astro-ph/9909414;
	J.R. Primack, M.A.K. Gross, astro-ph/0007165;
	J.R. Primack, astro-ph/0007187;
	J. Einasto, in Proc. of DARK2000, Heidelberg, Germany, July 10-15, 
	2000, Ed. H.V. Klapdor-Kleingrothaus, 
	{\it Springer, Heidelberg}, (2001).

\bibitem{Suz-Neutr2000} Y. Suzuki in Proc. of NEUTRINO2000, Sudbury, Canada, 
	June 2000, ed. A.B. McDonald et al. (2001).

\bibitem{Gonz00} M.C. Gonzalez-Garcia, M. Maltoni, C. Pe\~na-Garay, 
	J.W.F. Valle, {\it hep-ph/}{\bf 0009350}, 
	{\it Phys. Rev.} {\bf D63} (2001) 033005.

\bibitem{KK-AnnRep00} H.V. Klapdor-Kleingrothaus et al., 
	{\it Annual Report Gran Sasso 2000} (2001). 

\bibitem{KKAnnRep99-00} H.V. Klapdor-Kleingrothaus et al., MPI Heidelberg, 
	{\it Annual Report 1999-2000} (2001). 

%
\bibitem{KK1-NOW00} Talk on this conference H.V.Klapdor-Kleingrothaus 
	``GENIUS - A New Facility of Non-Accelerator Particle Physics''.

\bibitem{KK-NOON} H.V. Klapdor-Kleingrothaus in Proc. of NOON2000,  
	Tokyo, Dec. 2000, World Scientific, Singapore (2001).

\bibitem{Glash} H. Georgi and S.L. Glashow, {Phys. Rev.} {\it D 61} (2000) 
	097301.

\bibitem{Min97} H. Minakata and O. Yasuda, {\it Phys. Rev.} {\it D 56} (1997) 
	1692 and Minakata, {\it hep-ph/} {\bf 0004249}.

\bibitem{Yas-Bey00} O. Yasuda in Proc. of Beyond the Desert'99, ed. 
	by H.V. Klapdor-Kleingrothaus and I.V. Krivosheina, {\it IOP Bristol} 
	(2000) 223.

\bibitem{Ell99} J. Ellis and S. Lola, {\it Phys. Lett.} {\bf B 458} (1999) 
	310 and Preprint: {\it hep-ph/}{\bf 9904279}.

\bibitem{WeinLob-Neutr2000} C. Weinheimer in Proc, of NEUTRINO2000, Sudbury, 
	Canada, June 16 - June 21 (2000).

\bibitem{Teg-0008145} M. Tegmark, M. Zaldarriaga and A.J.S. Hamilton, 
	Preprint: {\it hep-ph/} {\bf 0008145}.

\bibitem{Weinh2000} Ch. Weinheimer in Proc. of NEUTRINO2000, Sudbury, Canada, 
	June 2000, ed. A.B. McDonald et al. (2001). 
 
\end{thebibliography}
\end{document}